\documentclass[a4paper,11pt]{article}
\usepackage{jcappub}
\usepackage{amsmath}
\usepackage{amssymb}
\usepackage{graphicx}

\title{Model Independent Prediction of the Spectral Index of Primordial Quantum Fluctuations}
 
 \author[a]{C\'esar G\'omez,} 
\emailAdd{cesar.gomez@uam.es}

\author[b,c]{Raul Jimenez} 
\emailAdd{raul.jimenez@icc.ub.edu}

\affiliation[a]{Instituto de F\'{i}sica Te\'orica UAM-CSIC, Universidad Aut\'onoma de Madrid, Cantoblanco, 28049 Madrid, Spain.}
\affiliation[b]{ICC, University of Barcelona, Marti i Franques 1, 08028 Barcelona, Spain.}
\affiliation[c]{ICREA, Pg. Lluis Companys 23, Barcelona, E-08010, Spain.}

\abstract{One of the most important achievements of inflationary cosmology is to predict a departure from scale invariance of the power spectrum for cosmological scalar perturbations. This tilt is understood as a consequence of a quasi de Sitter classical equation of state describing the inflationary dark energy dominated era. Here, following previous work, we find a departure of scale invariance for the quantum Fisher information associated to de Sitter vacuum for scalar quantum spectator modes. This gives rise to a purely quantum cosmological tilt with a well defined dependence on energy scale. This quantum tilt is imprinted, in a scale dependent energy uncertainty for the spectator modes. The effective quasi de Sitter description of this model independent energy uncertainty uniquely sets the effective quasi de Sitter parameters ( i.e., the small deviation from the cosmological constant equation of state) at all energy scales. In particular, in the slow-roll regime characterized by an almost constant $\epsilon$, the quantum Fisher --model independent-- prediction for the spectral index is $n_s=0.9672$. Moreover, the energy scale dependence of the quantum cosmological tilt implies the existence of a cosmological phase transition at energies higher than the CMB scale and of the order of $1Mpc^{-1}$ where the tilt goes from red into blue. This strongly suggest the existence of a pre-inflationary phase where the effective scalaron contributes to the spectral index as normal relativistic matter and where the corresponding growth of the power spectrum can result in dark matter in the form of small mass primordial black holes. The source and features of the quantum cosmological tilt leading to these predictions are determined by the entanglement features of the de Sitter $\alpha-$ vacuum states.}

\begin{document} 
\maketitle

\section{Introduction and brief summary}
In a series of recent papers \cite{GJ1,GJ2,GJ3,GJ4,GJ5} we have initiated a new approach to inflationary cosmology based on the notion of quantum Fisher information. It is fair to say that the idea has been, until now, mostly ignored by the community. Very likely this lack of interest is due to two basic reasons. First, the approach is based on notions as quantum Fisher information that are not very familiar among theoreticians creating a discouraging potential barrier for the reader. Secondly, we have probably been unable to convey in more pristine terms the physics underlying the whole approach. In this article, where we summarize some of the results and present new ones, we will try to correct both deficiencies.

Let us start saying a few words about quantum Fisher information (see Ref.~\cite{Paris} and references therein). The classical Fisher information, by contrast to the most popular Shannon information, intends to quantify, in the case of probability distributions depending on some statistical parameter, how much information we can extract about the actual value of the parameter. Classically this information is the width of the likelihood function when estimating some random parameters: the smaller the Hessian of the likelihood surface, the larger the value of the inverse of the Fisher matrix which is the uncertainty on the estimated parameter. In that sense, Fisher information is designed to account for the problem of estimating the value of a parameter about which we only have indirect and statistical information. In the same way that Shannon information has, as its quantum avatar, von Neumann entropy, in the case of Fisher we can define quantum Fisher information for families of quantum states (pure or mixed) depending on some parameter that by construction is not directly measurable i.e. that is not associated with any self adjoin physical observable. The first use of quantum Fisher information is to provide the metric measuring the {\it distinguishability  distance} between  states associated with different values of the external parameter. From a more physical point of view the most basic example of quantum Fisher information is obtained when we choose {\it time} as a external parameter. In this case the quantum Fisher is simply the {\it uncertainty of $G^2$ for $G$ the infinitesimal generator of time translations} i.e. quantum Fisher is just a way to measure $\Delta (E^2)$, where $E$ is the energy. In this case, the quantum estimation problem relative to the time parameter becomes simply the Heisenberg time energy uncertainty relation \cite{AA}.

Why can this story about quantum estimation of time be relevant to understand inflationary cosmology?

Inflationary cosmology \cite{Sta,Inf1,Inf2,Inf3,Inf4,Inf5} is the present paradigm for the cosmological description of the early Universe. The essence of inflationary models is to find solutions to the Einstein equations representing an expanding de Sitter phase (inflation). In addition, there is the need for a natural graceful exit taking place after a finite amount of time. This time needs to be long enough to account for the observable cosmic microwave background (CMB) and large scale structure (LSS) spectrum. An initial important step in that direction was taken by Starobinsky \cite{Sta} who noticed, while trying to solve the initial singularity problem, that Einstein's equations with an energy momentum tensor defined by the {\it trace anomaly} can give rise to an initial unstable de Sitter phase ending naturally after a finite amount of time. This initial model, where the physical parameters were determined by the matter content involved in the vacuum polarization responsible for the trace anomaly, was soon improved into a full fledged inflationary model \cite{Sta-Linde}. Indeed, the part of the trace anomaly that goes like $\Box R$ can be reabsorbed by adding a counter term $R^2$ to the Lagrangian with an arbitrary coefficient. This $R^2$ gravity has in its spectrum not only the graviton but also a scalar field that was identified with the original scalaron. However, the big secret of this model was only understood with the development of the theory of quantum cosmological perturbations \cite{CHM,MCH,Hawking,QF1,QF2,QF3,QF4}. The main observation in Ref.~\cite{MCH} was that the equation for the scalaron quantum fluctuations in the local de Sitter background \cite{CHM}, already in the linear approximation, leads to a power spectrum that is {\it not scale invariant}. The relevant parameter measuring the lack of scale invariance, the {\it tilt}, being roughly given by $\frac{M^2}{H^2}$ for $M$ parameterizing the scalaron mass. This non vanishing tilt leads to a time dependent power spectrum for scalar fluctuations and to a natural graceful exit whenever the amplitude of the fluctuations reach the values defining the background metric \cite{MCH}. In summary the theory of quantum fluctuations immediately leads to the basic observation that, provided a UV cutoff on the amplitude of the fluctuations, the existence of tilt naturally leads to a
graceful exit. 

But what is the relation between all this well known stuff and quantum Fisher information ?

Here comes our first basic observation. In the inflationary paradigm the equation for the scalar quantum fluctuations describes {\it time dependent quantum oscillators} with frequency $\omega^2(k,\eta)$ depending both on the comoving momentum and conformal time $\eta$. Formally this frequency contains three pieces. One is $k^2$ as it is the case for flat background. The second piece goes like $\frac{-2}{\eta^2}$ and describes the modification due to exact de Sitter. Finally, the last piece is the one that effectively accounts for the scalaron mass and goes roughly as $-\frac{M^2}{H^2} \frac{1}{\eta^2}$ \footnote{Note that $M^2$ enters into the dispersion relation with negative sign.}. The parameter $M$ that is {\it model dependent} and parametrized by the slow roll coefficients accounts for the classical design of the quasi de Sitter instability. In summary, the details of the quasi de Sitter slow roll are imprinted in the extra piece of the frequency.

Let us now come back to our discussion on quantum Fisher. Imagine yourself doing quantum field theory for a scalar spectator in pure de Sitter. In the $(k,\eta)$ hyperplane with $k$ representing the comoving momentum and $\eta$ the conformal time, we can define the hyperboloid ${\cal{H}}_{\Lambda}$ by the condition $k |\eta| H = \Lambda$. Thus each hyperboloid is associated with a given energy scale $\Lambda$. For each comoving momentum $k$ we define, using the standard Bogolioubov transform, the creation-annihilation operators $a_k(\Lambda)$ (for details see Ref.~\cite{GJ4} and references therein) and the corresponding $|\Lambda\rangle$ vacuum. In essence, this state represents the vacuum for modes of comoving momentum $k$ and energy scale $\Lambda$. A peculiar property of quantum field theory in de Sitter is that for a given momentum $k$ the energy scale depends on the conformal time. This leads to some difficulties at the time of defining effective field theory in Wilsonian terms. We will not touch this issue in this article. The key ingredient of our approach is the computation of the quantum Fisher information for the state $|\Lambda\rangle$ defined relative to a Fock like basis. This Fock basis is defined using the standard euclidean Chernikov-Tagirov~\cite{CT}/Bunch-Davies~\cite{BD} vacuum. In this basis the $|\Lambda\rangle$ vacuum is defined, as usual, relative to the corresponding Bogoloubov transformed oscillators. Physically, this information is telling us the maximal information we can  extract, about the value of $\Lambda$ by performing measurements on the state $|\Lambda\rangle$. Note that by the definition of de Sitter energy scale this information is telling us about the quantum uncertainty in the conformal time and consequently in the physical energy scale. 

The main result we have derived in Ref.~\cite{GJ4} is the renormalization group equation for the Fisher information. In reduced notation, and denoting $F(\Lambda)$ the quantum Fisher information modulo, the conformal piece (see section 3 for details and definitions), the basic result is
\begin{equation}\label{main}
\Lambda\frac{\partial}{\partial \Lambda} F = \alpha_F F
\end{equation}
or equivalently $\frac{d \ln F}{d \ln \Lambda} = \alpha_F(\Lambda)$. This means that quantum Fisher is not scale invariant and identifies the quantum tilt $\alpha_F$ as well as its dependence on the energy scale. It is important to stress that this anomalous scale transformation of quantum Fisher has as its origin the specific form of the quantum phases introduced through the Bogolioubov transform defining the state $|\Lambda\rangle$. It is well known that the state $|\Lambda\rangle$ is a very entangled squeezed state (see \cite{SQU1} see also \cite{GJ4} and references therein). The quantum Fisher as well as the shape of the quantum tilt strongly depends on the entanglement features \footnote{In order to avoid some potential missunderstandings it is important to make some basic distinctions. For the squeezed state that is strongly entangled we can define von Neumann density matrix integrating over one of the partner modes defining the squeezed state and to evaluate the corresponding entanglement entropy. Moreover we can consider families of von Neumann density matrices depending on de Sitter energy scale and to evaluate the relative entanglement entropy between them and its relation with a Fisher information metric on the space of mixed states. In this note we focus on the quantum Fisher of the {\it pure state}.} 

After these preliminaries, we are in a good position to establish our main claim about how  to establish a correspondence between the tilt of the spectral index of scalar fluctuation in quasi de Sitter and the value of the quantum Fisher when the estimated parameter is time. From Fisher information we can derive the uncertainty of the energy square on the state $\Lambda$; let us denote it $\Delta(E^2)(k\eta)$ and we can compare it with the corresponding energy uncertainty defined in the quasi de Sitter set up $\delta(E^2)(k,\eta)$, where this second energy depends on the slow roll parameters. Both transform anomalously with respect to changes of energy scale. In the case of Fisher $\Delta(E^2)$ the tilt is given by $\alpha_F$, while in the quasi de Sitter case is given by the cosmological spectral index. To put both in correspondence let us identify both quantities. This defines a self consistency constraint between the slow roll parameter $2\delta= 1-n_s$ and the value of the Fisher quantum tilt $\alpha_F$ that uniquely fix the value of the slow roll parameter, 

\begin{equation}\label{self}
\boxed{\frac{6\delta + 2\delta^2}{\alpha_F^{-1}(4\delta)} =1},
\end{equation}

where $\alpha_F^{-1}$ is the inverse of $F$. Solving this simple equation (which is derived in detail in the sections below) for $\delta$ for the Fisher tilt $\alpha_F$ fixes the value of the cosmological spectral index $1-n_s$ in the slow roll to be
\begin{equation}
\boxed{1-n_s= 0.0328 }
\end{equation}

This result that magically fits the one from Planck18~\cite{Planck18} $1-n_s= 0.0348 \pm 0.0042$, strongly indicates that the cosmological spectral index is fully determined by identifying the quantum Fisher energy uncertainty with the energy uncertainty we associate with quasi de Sitter slow roll, i.e. with the quasi de Sitter piece of $\omega^2(k,\eta)$ simply defined as $\omega^2-k^2-\frac{2}{\eta^2}$. Note that in order to get the value of $1-n_s$ we don't have any model dependent freedom. This number is fully determined by the quantum Fisher information. The nice agreement with experiment implies that in the CMB scales the slow roll approximation is extremely good.

Before entering into the technical derivation of (\ref{self}) let us summarize for the benefit of the reader the main steps of our construction.

\begin{enumerate}
\item First, we define the quasi de Sitter contribution to the oscillator energy $\Delta E^2(k,\eta,\delta)$ as $E^2(k,\eta,\delta)-E^2(k,\eta,\delta=0)$, where $E^2(k,\eta,\delta) \equiv \frac{\omega^2(k,\eta,\delta)}{a^2(\delta)}$ with $a(\delta,\eta) = \frac{1}{H_0\eta} \frac{1}{(k_0\eta)^{\delta}}$. Here $k_0$ is a pivot scale defined by the condition $a(\delta,\eta_0) = \frac{1}{H_0 \eta_0}$ ( for more details see section 2 ). Note that this quantity depends on $k,\eta,\delta$ and is zero for $\delta=0$. 

\item Now, you put yourself in pure de Sitter and identify {\it the quantum variance}
$\delta(E^2(k,\eta))$ defined by the quantum Fisher information evaluated at $(k,\eta)$. The derivation of this quantity is the main result of section 3. This quantity only depends on pure de Sitter data. The key point that substantiate our derivation is that this pure de Sitter quantity depends on the quantum tilt $\alpha_F$ defined by the anomalous scale dependence of the pure de Sitter quantum Fisher information ( see (\ref{main})). 

\item Finally we ask ourselves about what quasi de Sitter background can lead to exactly the same value that the pure de Sitter quantum Fisher result. The answer to this question is determined by the equation:
\begin{equation}\label{eq:two}
 \Delta E^2(k,\eta,\delta) = \delta(E^2(k,\eta))
\end{equation}
\end{enumerate}

This identification defines a very non trivial constraint. First of all $\Delta E^2(k,\eta,\delta)$ was defined in the slow roll parametrization of quasi de Sitter. Secondly $\delta(E^2(k,\eta))$ was defined in pure de Sitter. Both quantities define a correction to the oscillator frequency of the same order in $\hbar$. Moreover both quantities include the effect of anomalous scale invariance. In the case of $\Delta E^2(k,\eta,\delta)$ this effect is parametrized by $\delta$ while in the case of $\delta(E^2(k,\eta))$ this effect is encoded in $\alpha_F$. Now note that the solution to equation (\ref{eq:two}) with data $\alpha_F$ selects a slow roll parameter very non trivially. First of all the quantum anomalous scale invariance selects a function $k\eta(\delta)$ through the relation, discussed in section 3,  $\alpha_F(k\eta)=4\delta$. This defines the matching between the scale dependence set by the slow roll parameter $\delta$ with the one defined, model independently, by the quantum Fisher tilt $\alpha_F$. The important thing now is that in order to achieve (\ref{eq:two}) we need an extra condition that combined with the matching for the scale dependence described above leads to the main equation (\ref{self}). Solving this equation allows you to identify a unique value of $\delta$ and consequently of the spectral index. 

The careful reader can find this claim paradoxical in the following sense. The slow roll piece of $\omega^2$ is fully determined by the form of the classical slow roll potential that we can model at will. Nevertheless, the quantum Fisher is in origin quantum and comes from the phases of the state $|\Lambda\rangle$, consequently we have not any freedom to change its value. The big claim we are making in this article is that the quantum Fisher effect {\it fully accounts for the observed value of the cosmological tilt}. 

Superficially the claimed {\it quantumness} of the tilt seems to enter into conflict with the general picture of the classicalization in time of the fluctuations. Indeed, the problem of how the quantum coherence of the de Sitter state affects the correlators we normally use to compare with observations has been extensively studied (see for instance Ref.~\cite{entropy1,entropy2,entropy3,entropy4}). The basic point is that the observational effects of the phases responsible for the quantum coherence decrease exponentially with time as $e^{-{\cal N}}$ for ${\cal N}$ the number of e-foldings. This puts, in particular, severe constraints to cosmological Bell like experiments~\cite{GreenPorto}. Hence the obvious question is: Why this classicalization of fluctuations is not in contradiction with the quantum Fisher approach? How can we claim that we can extract the observable tilt from quantum effects of the de Sitter state when the quantum coherence of these effects is exponentially suppressed in the limit of deep IR scales corresponding to $k\eta << 1$ ? The essence of the answer is simply the following. Using quantum Fisher information we discover a quantum cosmological tilt for all energy scales. In the deep IR i.e. for $k\eta <<1$ the suppressed contribution of quantum phases will lead in the limit $k\eta=0$ to a vanishing quantum Fisher and a vanishing tilt. This effective classicalization is however not the end of the story. Indeed the quantum Fisher and its associated tilt is not vanishing  for small values of the energy scale. Now the possibility that quantum Fisher effects account for the cosmological tilt is determined by the solution to (\ref{self}) that depends on the particular energy dependence of the quantum tilt away from the very IR limit $k\eta=0$. It is the non trivial dependence of the quantum tilt away from the deep IR limit what sets the observable value of $1-n_s$. In more plain words: the quantum Fisher in pure de Sitter accounts for the effect we simulate classically using quasi de Sitter slow roll. This makes the spectral index a fundamental label of our Universe. More precisely it can be interpreted as the basic label of a {\it de Sitter universality class}. 

In the quantum Fisher approach presented in this article, we reduce the study of quantumness to the de Sitter "vacuum" for the scalar spectator. Other quantum effects due to the pure gravitational constituency of the metric background have been recently considered in \cite{Gia1,Gia2,Gia3,Gia4}. These effects go as $1/N_{GH}$ for $N_{GH}$ the de Sitter entropy and define quantum back reaction corrections on the top of the ones we discuss here.

\begin{figure}
\centering
\includegraphics[width=.9\columnwidth]{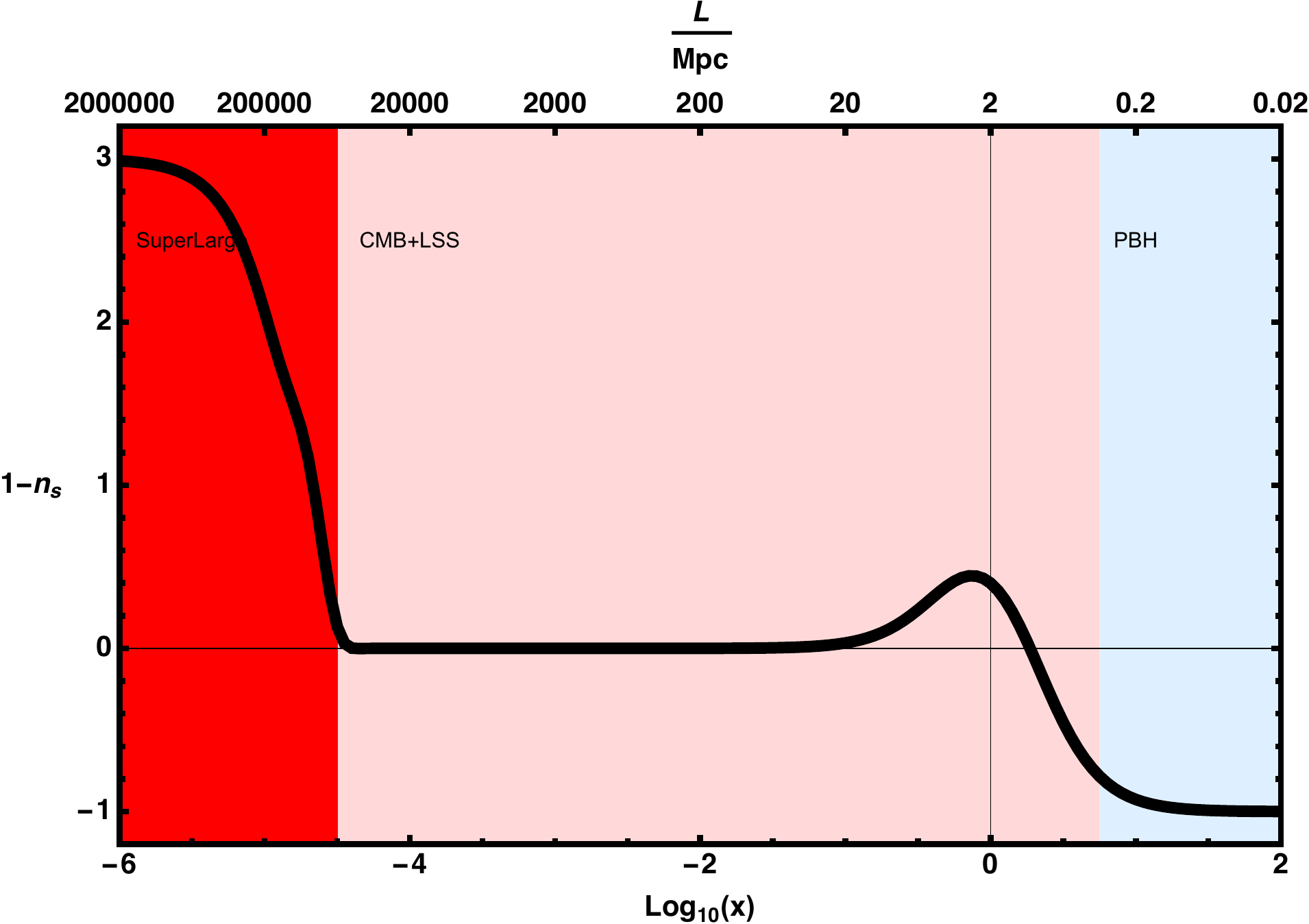}
\caption{The quantum cosmological tilt as a function of energy scale as determined by the quantum Fisher. For values of $|k \eta| < 0 $ the end of inflation is determined by the IR cutoff on the number of particles $n$ in the sum (3.2). In this plot, for illustrative purposes, we set $n=10^9$. The sensibility of the quantum cosmological tilt on this IR cutoff is discussed in the appendix showing that only affects the effective duration of inflation but not the predicted value for $(1-n_s)$. The different shaded regions correspond to different scales. The blue region is where the dark matter forms and the spectral index is blue. A detailed explanation of each region is given in the appendix.}
\label{fig:qtilt}
\end{figure}

A bonus of the Fisher approach is that it gives us a tilt with a complete dependence on the energy scale (see Fig.~\ref{fig:qtilt})\footnote{In this figure we represent the quantum tilt in terms of the energy scale. The sensibility of this numerical result on the number of created pairs is discussed in the appendix.}.  Several things about this dependence are worth to be highlighted. First of all we see an almost scale invariant region corresponding to the inflationary period. As discussed in the appendix the extension of this region depends on an effective IR cutoff. Secondly, we observe a change of phase {\it from red tilted into blue tilted at high momentum}. The change in the tilt reflects an effective change in the nature of the scalaron. Indeed the tilt can be parameterized in terms of a running effective scalaron mass that changes sign in the blue tilted phase. This blue tilted phase can naturally account for the dark matter, in the form of primordial black holes, needed to amplify gravitationally the primordial quantum fluctuations (see e.g. Ref.~\cite{BH} for a recent review and references therein). 

\section{Standard theory of quantum cosmological fluctuations}
The fundamental ingredient to develop the theory of gauge invariant quantum cosmological scalar fluctuations is the Chibisov-Mukhanov \cite{CHM} equation \footnote{For the derivation see appendix in \cite{CHM}}
\begin{equation}
\phi^{''} - \Delta \phi -\frac{z^{''}}{z}\phi =0
\end{equation}
with $z= a\sqrt{\epsilon}$ and $\epsilon$ the standard first slow roll parameter. For the Fourier transform components we get 
\begin{equation}\label{eq:one}
\phi_{k}^{''} + ( k^2 - \frac{z^{''}}{z}) \phi_{k} =0
\end{equation}
leading to an effective frequency $\omega^2(k,\eta) = (k^2-\frac{z^{''}}{z})(k,\eta)$, where $\eta$ is the conformal time. Using
\begin{equation}\label{zz}
\frac{z^{''}}{z}= \frac{a^{''}}{a} + \frac{a^{'}}{a}\frac{\epsilon^{'}}{\epsilon} +\frac{1}{2} \frac{\epsilon^{''}}{\epsilon} - \frac{1}{4} \frac{(\epsilon^{'})^2}{\epsilon^2}
\end{equation}
we can define the simplest {\it slow roll} approximation by the condition $\epsilon$ constant but different from zero and 
\begin{equation}
\frac{z^{''}}{z}= \frac{\beta(\beta+1)}{\eta^2}
\end{equation}
This corresponds to use a scale factor
\begin{equation}\label{metric}
a= \frac{1}{H_0\eta} \frac{1}{(k_0\eta)^{\delta}}
\end{equation}
where we introduce, for dimensional reasons, a reference momentum $k_0$. Independently of the value of $k_0$ we get
\begin{equation}
\frac{a^{''}}{a} = \frac{\beta(\beta+1)}{\eta^2}
\end{equation}
with $\beta = -2-\delta$. The Hubble parameter defined as $H=\frac{\dot a}{a}$ is given by
\begin{equation}\label{Hubble}
H(\eta) =(\delta+1)H_0(k_0\eta)^{\delta}
\end{equation}
leading to a constant value of the slow roll parameter $\epsilon= -\frac{\dot H}{H^2}$ given by
\begin{equation}
\epsilon =\frac{\delta}{\delta+1}
\end{equation}
Let us now consider equation (\ref{eq:one}) in this slow roll limit for a given $\beta$. In the {\it long wave limit} $k\eta <<1$ the equation reduces to a Bessel equation \footnote{In the case of constant $\epsilon$ the exact solution is given in terms of Hankel functions.}. The solution is given by
\begin{equation}
k^3 |\phi_k(\eta)|^2 \sim \frac{1}{\eta^2} (k\eta)^{-2\delta}
\end{equation}
Thus defining the power spectrum as $P(k,\eta) \equiv \frac{1}{\epsilon a^2 M_P^2} k^3|\phi_k|^2$ we get in this long wave length limit \cite{MCH} and using (\ref{metric})
\begin{equation}\label{power}
\boxed{P(k,\eta) = \frac{H_0^2}{\epsilon M_P^2} (\frac{k_0}{k})^{1-n_s}}
\end{equation}
with $(1-n_s) = 2\delta$.
Note that for this definition of the power spectrum:
\begin{itemize}
\item $k_0$ plays the role of pivot scale
\item We work in the slow roll approximation with $\epsilon$ constant
\item The tilt $1-n_s$ is computed in the limit $k\eta <<1$
\item $H_0$ is the initial value at time $\eta_0=\frac{1}{k_0}$. More precisely
$H_0=\frac{H(\eta_0)}{(1+\delta)}$.
\end{itemize}
If we choose for $k_0$ the CMB scale $k_{\rm CMB}$ then equation (\ref{power}) sets the power spectrum for $H_0$ the value of Hubble at $\eta = \frac{1}{k_{\rm CMB}}$ i.e. at horizon exit. This is the standard prescription where you use the Hubble at horizon exit time of CMB modes and approximate the tilt using constant $\epsilon$ and the solution to Bessel in the long wave limit. To improve this approximation you need to consider the more general case with $\epsilon$ depending on time and to solve exactly in that regime equation (\ref{eq:one}).
\subsection{Energy and scale invariance}

The physical energy of a mode of comoving momentum $k$ at conformal time $\eta$ is given, in the slow roll regime, by 
\begin{equation}
E^2(k,\eta) = \frac{1}{a^2}(k^2-\frac{\beta(\beta+1)}{\eta^2})
\end{equation}

Let us define variation of the energy relative to the pure de Sitter case. This is given by
\begin{equation}\label{sruncertainty}
\boxed{\Delta E^2(k,\eta) = (3\delta + \delta^2)H_0^2(k_0\eta)^{2\delta}}
\end{equation}
In the limit of $\delta <<1$ we simply get
\begin{equation}
\Delta E^2(k,\eta) \sim 3\delta H(\eta)^2
\end{equation}
The lack of scale invariance can be nicely expressed in the form  of a {\it renormalization group} equation, namely
\begin{equation}
\boxed{k_0\frac{\partial}{\partial k_0} \Delta E^2 = 2\delta \Delta E^2}
\end{equation}
It is an interesting exercise to evaluate $\Delta E^2$ for the Starobinsky model \cite{Sta} in terms of the {\it scalaron} mass $M$ as defined in \cite{MCH}. In this case $2\delta= \frac{2M^2}{3H^2}$ and we get
\begin{equation}
\Delta E^2= M^2 (k_0\eta)^{2\delta}
\end{equation}
In other words, the {\it scalaron mass} $M^2$ is simply $\Delta E^2$ evaluated at the reference time $\eta_0 =\frac{1}{k_0}$. Moreover the former expression leads to a running scalaron mass
\begin{equation}
M^2(\eta) = M^2 (k_0\eta)^{2\delta}
\end{equation}
Thus defining $\eta_0$ as initial time the bound on the time life of quasi de Sitter can be derived setting the final time $\eta_f$ by the bound
$M^2(\eta_f) = M_P^2$ that leads to
\begin{equation}\label{bound}
\delta t = \frac{1}{H} \frac{3H^2}{2M^2} \ln(\frac{M_P^2}{M^2})
\end{equation}
that is the estimation of quasi de Sitter time in Ref.~\cite{MCH}.

As a comment note that in the pure de Sitter limit i.e. $M=0$ we get $\delta t =\infty$ and $\Delta E^2 = 0$. This is in essence a manifestation of energy time uncertainty relation, namely {\it for eternal de Sitter $\Delta E^2$ is zero}.\\

In summary we observe the natural set of connections:\\

Lack of conformality $\Leftrightarrow$ Tilt $\Leftrightarrow$ graceful exit $\Leftrightarrow$ energy "uncertainty" $\Delta E^2 \neq 0$

\subsection{Connection to quantum breaking}
In the approach developed in Ref.~\cite{Gia1,Gia2,Gia3,Gia4} the life time of de Sitter based on quantum breaking was estimated to be
\begin{equation}
\delta t \sim \frac{1}{H}\frac{M_P^2}{H^2}
\end{equation}
It is instructive to compare this time with the bound (\ref{bound}). Formally the quantum breaking corresponds to a scalaron mass 
\begin{equation}
\boxed{M= \frac{M_P}{N_{GH}}}
\end{equation}
with $N_{GH} \equiv \frac{M_P^2}{H^2}$ the de Sitter Gibbons-Hawking entropy. The simplest interpretation is that pure quantum breaking effects in de Sitter lead naturally to an effective scalaron mass of order $1/N_{GH}$. This also highlights that eternal de Sitter corresponds to the $N_{GH}=\infty$ limit. This quantum scalaron mass has its origin in the quantum constituency of de Sitter and leads to a tilt of order $1/N_{GH}$ \footnote{In order to avoid any confusion note that although a finite value of de Sitter entropy was originally obtained using an eternal de Sitter metric the constituency representation of de Sitter where we look for a coherent state -- defined relative to Minkowski vacuum-- with the expectation value of the metric reproducing classical de Sitter geometry has an occupation number of the order $N_{GH}$.Hence constituency effects are always order $\frac{1}{N_{GH}}$. }

As already discussed in the introduction, this $1/N_{GH}$ tilt is not the one we discover using the quantum Fisher for the scalar spectator in de Sitter. In the latter case the Fisher quantum tilt extracts the anomalous scale transformations relative to changes of energy scale for the dS vacuum for the spectator modes. We will use the next section to derive the quantum Fisher tilt and present its connection with the cosmological tilt for quantum scalar fluctuations.

\section{The quasi de Sitter-Fisher correspondence}
\subsection{de Sitter kinematics and Fisher information}
\begin{figure}
\centering
\includegraphics[width=.9\columnwidth]{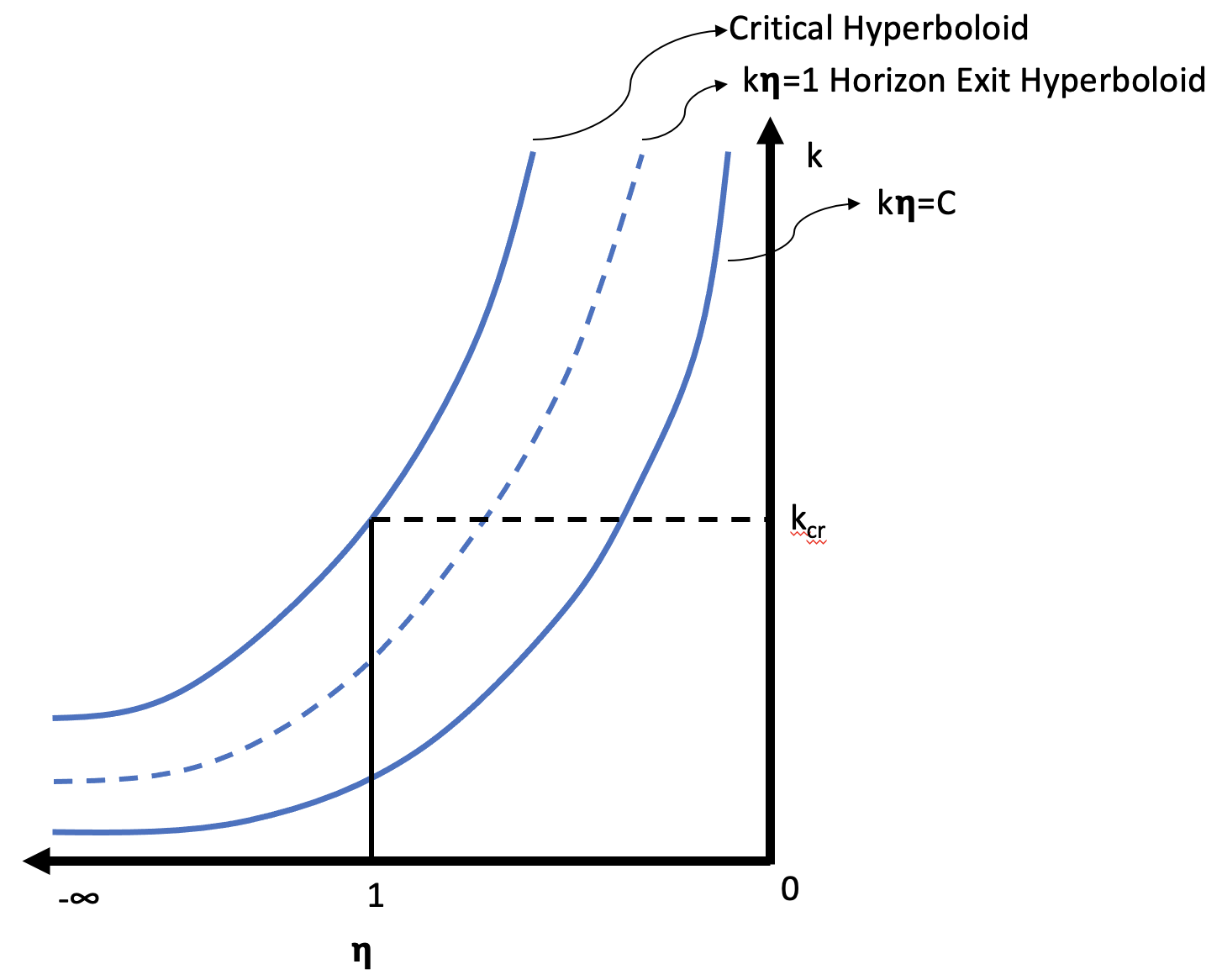} 
\caption{The plane $k,\eta$ foliated in hyperboloids representing the same energy scale. The horizon exit hyperboloid is $k\eta=1$.}
\label{fig:hyper}
\end{figure}

Observables in inflationary cosmology are functions of the comoving momentum $k$ and the conformal time $\eta$. Conformal time is defined in the interval $[-\infty,0]$ covering the full physical time interval $[-\infty, +\infty]$. It is convenient to foliate the $k,\eta$ hyperplane into hyperboloids defined by the condition $k\eta=C$, for $C$ arbitrary constant (see Fig.~\ref{fig:hyper}). In pure de Sitter we can define the energy scale $\Lambda= -k\eta H$ so these hyperboloids define equal energy hypersurfaces. 
 
For a scalar spectator we can define for each couple $(k,\Lambda)$ a dS {\it "vacuum" state} as
\begin{equation}\label{state}
|k,\Lambda\rangle = \sum_n c(\Lambda,n) e^{i n \Phi(\Lambda)}|n_k,n_{-k}\rangle
\end{equation}
where we sum over all integers and  
\begin{equation}
c(\Lambda,n)=C \tanh({r(\Lambda)})^{n}
\end{equation}
 with $C$ a normalization constant, 
\begin{equation} 
 r(\Lambda) = - \sinh^{-1}(\frac{H}{2\Lambda})
\end{equation} 
and
\begin{equation}
\Phi(\Lambda) = -\frac{\pi}{4} -\frac{1}{2} \tan^{-1}(\frac{H}{2\Lambda}). 
\end{equation} 
   
Using the definition of $\Lambda$ this state represent the dS vacuum for mode $k$ at conformal time $\eta= \frac{\Lambda}{kH}$. Thinking of $\Lambda$ as statistical parameter the dependence of the state on $\Lambda$ simply reflects the conformal time dependence for the mode $k$. 

Already at this level we can define the quantum Fisher information as
\begin{equation}
F(k,\Lambda) = k^2( \sum_n c(\Lambda,n)^2 (\frac{\partial \Phi(\Lambda,n)}{\partial \eta})^2 - (\sum_n c(\Lambda,n)^2 (\frac{\partial \Phi(\Lambda,n)}{\partial \eta}))^2)
\label{eq:eqfisher}
\end{equation}
which is, note the overall factor $k^2$, the quantum Fisher information defined relative to the conformal time $\eta$. For pure de Sitter we get
\begin{equation}
F(k,\Lambda) = \frac{1}{\eta^2(k\eta)^6} (k\eta)^{\alpha_F(k\eta)}
\end{equation}
where $\alpha_F(k\eta)$ is the {\it quantum Fisher tilt} that has a very precise dependence on $(k\eta)$ (see Fig.~\ref{fig:qtilt}). In order to unravel the physical meaning of this quantum tilt we need to understand how this quantum Fisher tilt enters into the quantum uncertainty of physical energy.~\footnote{See the appendix for a detailed description of the energy dependence of $\alpha_F(k \eta)$.}

First of all, note that for each value of the energy scale $\Lambda$ we can define a family of quantum states corresponding to use the different values of $k,\eta$ on the hyperboloid defined by $\Lambda$. In reality, we have {\it two identical hyperboloids:} one for the mode $k$ and another for the mode $-k$ entering into the definition of the state (\ref{state}). Thus after taking into account the integration factor over the two hyperboloids we get
\begin{equation}
F= \frac{1}{\eta^2}(k\eta)^{\alpha_F(k\eta)}
\end{equation}
This leads to a quantum uncertainty on $k$ at energy scale $\Lambda$ given by the square root of the quantum Fisher function, namely
\begin{equation}
\delta(k) = \frac{1}{\eta} (k\eta)^{\alpha_F/2}
\end{equation}
Now we can define, at first order, $\delta E_{F}^2$ at energy scale $\Lambda$ as
\begin{equation}\label{Funcertainty}
\boxed{\delta E_{F}^2 (k\eta) = \frac{k\delta(k)}{2 a^2} = (\frac{k\eta}{2}) H^2 (k\eta)^{\alpha_F/2}}
\end{equation}
We should interpret this quantity as the quantum energy square uncertainty for spectator modes living on the hyperboloid defined by $\Lambda$. 
\subsection{The correspondence}
In order to define the quasi de Sitter-Fisher correspondence we need to compare the quasi de Sitter energy uncertainty given in (\ref{sruncertainty}) and Fisher quantum uncertainty given in (\ref{Funcertainty}). Formally:
\begin{equation}
\boxed{\delta E_{F}^2(k\eta) \Leftrightarrow \Delta E^2(k,\eta)}
\end{equation}
where $\Delta E^2(k,\eta)$ is the quasi de Sitter expression. To define the exact correspondence in the slow roll limit we choose $H$ used in the Fisher construction to be $H_0$ and we put ourselves in the energy scale implicitly defined by $\hat{(k\eta)}$ where
\begin{equation}\label{scale}
\alpha_F(\hat{(k\eta)}) = 4\delta
\end{equation}
with $\delta$ defined in (\ref{metric}). In these conditions we get the slow roll-Fisher correspondence at the scale defined by (\ref{scale})
\begin{equation}\label{correspondence}
\boxed{\Delta E^2 =\delta E_F^2 \frac{6\delta + 2\delta^2}{\hat{(k\eta)}} (\frac{k_0}{k})^{2\delta}}
\end{equation}
This equation has a very nice meaning. Indeed, it gives us the renormalization group running of the physical energy uncertainty in terms of the pivot scale and uses (model independent) quantum Fisher information to set the value at a pivot scale defined by (\ref{scale}).
\subsection{Fisher derivation of spectral index}
Recall that in the slow roll approximation we have $2\delta= 1-n_s$. Let us now derive this spectral index from imposing the correspondence $\Delta E^2 =\delta E_F^2$. From (\ref{correspondence}) this implies, for $k=k_0$,
\begin{equation}
\frac{6\delta + 2\delta^2}{\hat{(k\eta)}} =1
\end{equation}
Thus, {\it given as data the quantum Fisher tilt function $\alpha_F(k\eta)$} the spectral index in slow roll is determined by the equation
\begin{equation}\label{spectral}
\boxed{\frac{6\delta + 2\delta^2}{\alpha_F^{-1}(4\delta)} =1}
\end{equation}
Note that this equation is setting the value of the slow roll spectral index by imposing that:\\

{\it The quasi de Sitter energy uncertainty is equal to the quantum Fisher energy uncertainty.}\\

Note that the only data entering into this equation is the quantum Fisher tilt $\alpha_F(k\eta)$ that is completely model independent. Since we are interested in extracting $(1-n_s)$ at slow roll we impose as a self consistency condition the equality in this regime between the quasi de Sitter energy uncertainty and the quantum Fisher energy uncertainty. This a priori leaves the value of the slow roll parameter $\delta$ completely determined by $\alpha_F$. It is easy to solve the self consistency equation (\ref{spectral}). Surprisingly enough the result is
\begin{equation}
\boxed{1-n_s= 0.0328}
\end{equation}
i.e. $n_s=0.9672$.  This apparently magic result can induce the skeptic reader to suspect that we are cheating and that we are already introducing the information we want to extract. It is easy to convince yourself that the value of $1-n_s$ could be very different from the correct result for a different value of the quantum Fisher tilt $\alpha_F(k\eta)$. The key point is that this quantum tilt is completely independent of any quasi de Sitter modeling.

Coming back to our former discussion on quantum breaking, we should add to $1-n_s$ the quantum constituency contribution \cite{Gia1,Gia2,Gia3} $1/N_{GH}$ that is in general, for realistic values of $H$, negligible.

\subsection{Master equation}
For generic scale dependence of $\epsilon$ we have for $\frac{z^{''}}{z}$ the complicated function of the first and second derivative of $\epsilon$ given in (\ref{zz}). The exact tilt in this case will require to solve equation (\ref{eq:one}) exactly. The quasi de Sitter-Fisher correspondence described above leads to an implicit master equation relating the function of $\epsilon$ defined by the exact value of $\frac{z^{''}}{z}$ and the corresponding exact tilt that we will represent as $(1-n_s)(\epsilon)$. Let us define\footnote{Note that $A(\epsilon) = 2H^2a^2\eta^2(1-\frac{\epsilon}{2} - \frac{3}{4}\frac{d\ln \epsilon}{dN} + \frac{1}{4}\epsilon \frac{d\ln \epsilon}{dN} + \frac{1}{8}(\frac{d\ln \epsilon}{dN})^2 + \frac{1}{4}\frac{d^2 \ln \epsilon}{d^2N})-2$ with $dN=-aHd\eta$. In what follows we will assume that $\epsilon$ is an arbitrary power law function of $k\eta$.}
\begin{equation}
    A(\epsilon) \equiv \eta^2 \frac{z^{''}}{z} -2
\end{equation}
The master equation defining the quasi de Sitter-Fisher correspondence is formally given in terms of the exact Fisher tilt $\alpha_F$ by
\begin{equation}\label{Master}
  \boxed{\alpha_F^{-1}(2(1-n_s)(\epsilon)) = A(\epsilon)}
\end{equation}
    In the former paragraph we have considered this equation in the slow roll limit with $\epsilon$ constant but non vanishing. In this case $A(\epsilon) = 6\delta+2\delta^2$ and $2(1-n_s)(\epsilon) = 4\delta$ \footnote{Recall that this is the exact Bessel parametrization in the limit of $\epsilon$ constant.}. In this limit the master equation becomes equation~\ref{spectral} with the advertised solution $1-n_s= 0.0328$.We can describe in words the meaning of the master equation as follows:\\
    \begin{center}
    \fbox{$\alpha_F$ (quasi de Sitter energy difference) = spectral index} \\
    \end{center}
    i.e. the quantum cosmological Fisher tilt maps the deviation from de Sitter of the spectator dispersion relation, at a given scale, into the spectral index at that scale.

    However the master equation allows us to extract  information about the running of the exact spectral index beyond the slow roll regime. The physics implications of such a running dictated by the quasi de Sitter-Fisher correspondence will be discussed in next section.

\section{The non perturbative information of the quantum Fisher tilt: glimpses of a theory of dark matter}

In standard inflation the spectral index is computed using the solutions to equation (\ref{eq:one}) in the approximation where the slow roll parameters are almost constant. Deviations from this approximation require a model dependent specification of the slow roll parameters at different energy scales. The key bonus of the Fisher approach is that it provides a well defined dependence of the quantum tilt $\alpha_F(k\eta)$ on energy scale. In this section we will try to extract some glimps on how a full fledged cosmology can fit the shape of the quantum tilt.

\subsection{The critical hyperboloid: the effective scalaron mass}

From the explicit expression of $\alpha_F(k\eta)$ discussed in Ref.~\cite{GJ2} we discover a critical hyperboloid defined by
\begin{equation}
-(k\eta)_{cr} = 1.87
\end{equation}
where the quantum tilt {\it goes from red into blue}. For a fixed value of conformal time $\eta$ we can define $k_{exit} \sim \frac{1}{\eta}$ and the critical $k_{cr}$ defined by $k_{cr}\eta = 1.87$. This momentum is much larger than the horizon exit momentum. Thus this larger momentum will exit the horizon at a later conformal time.

What is the meaning of this cosmological phase transition?

The simplest way to address this question is using our master equation (\ref{Master}). Formally this equation is telling us that $\alpha_F$ at the effective scalaron mass is giving us the value at that scale of the spectral index $(1-n_s)$. In a very rough approximation. we can define the effective scalaron mass $M_{\rm eff}(k\eta)$ at energy scale $k\eta$ in terms of $\alpha_F$ as follows
\begin{equation}\label{Tilt}
    \alpha_F(k\eta)/2 = 3\sqrt{1 + \frac{4M^2_{\rm eff}(k\eta)}{9H^2}} -3
\end{equation}
Once we use this effective scalaron fit of the quantum tilt we inmediately observe that when crossing the critical hyperboloid we transit from the scalaron describing the inflationary period corresponding to the sign $+$ inside the square root to a pre-inflationary period where the sign changes into $-$ (see also Ref.~\cite{GJ5}). This means that for these large energies, larger than the critical energy determined by the critical hyperboloid, the effective scalaron mass square changes sign. This is extremely interesting since it indicates that the quantum effects imprinted in the quantum tilt are defining a phase transition to a phase where the Universe is contracting in time until reaching the critical time where the Universe enters into the standard inflationary phase. In this pre-inflationary phase the effective scalaron mass is entering in the dispersion relation as a normal matter mass effect.

\begin{figure}
\centering
\includegraphics[width=.9\columnwidth]{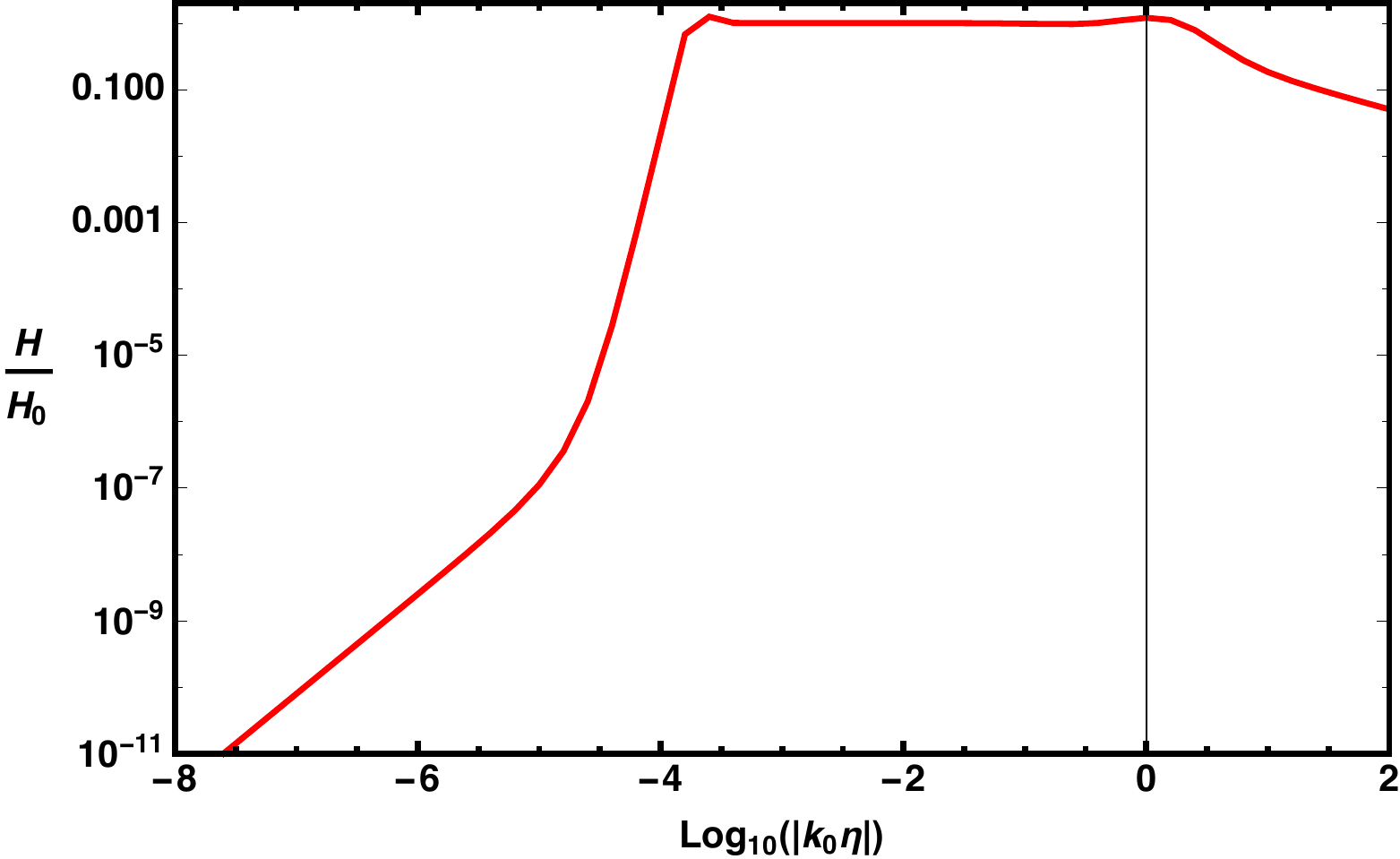}
\caption{The Hubble parameter as a function of scale for $k_0=1$. As in previous figures we have adopted $n=10^9$ for illustration purposes to set the end of inflation. Note that at very large and small scales, the Hubble parameter approaches zero. For the largest scales, the ones entering the horizon today, the expansion resembles that of a quintessence field after the FLRW phase.}
\label{fig:hubble}
\end{figure}

In simpler words; the blue tilt regime is dominated by a form of matter that is responsible for the change of sign in (\ref{Tilt}). An interesting interpretation for cosmology of our derived value for the quantum tilt is to use the Hubble defined by (\ref{Hubble}), which contains the information on  $\alpha_F$.
The result is depicted in Fig.~\ref{fig:hubble}. This figure indicates a pre-inflationary contracting phase and an end of inflation depending on the IR cutoff on the contributing number of pairs (see appendix). A pre-inflationary contracting phase is very much reminiscent of bouncing cosmological models \cite{BC}. Here we reduce ourselves to suggest this simple interpretation of the quantum cosmological tilt. It is well known that bouncing solutions enter into conflict with the null energy condition NEC \cite{NEC}. In our approach the change red to blue of the tilt is a pure quantum effect and we don't have an effective way to describe this change of phase semiclassically, thus we don't have anything to say on the interesting problem of violations of the NEC.

\subsection{Is the pre-inflationary high energy phase the source of dark matter?}

Although the primordial quantum fluctuations play the natural role of seeds of the large scale structures, the gravitational amplification of these seeds requires the presence of dark matter that interacts extremely weakly with the electro-magnetic force.
 In other words, we need (cold) matter but cannot be visible to do the job of galaxy formation properly. A very nice and natural suggestion on the nature of this matter are the so called primordial black holes. The advantage of this candidate is that they have a pure gravitational origin if we manage to create quantum fluctuations intense enough to create small black holes in a very concrete range of masses, bounded by Hawking evaporation, to allow  them to survive longer than Hubble time. To have such strong primordial fluctuations we need as a necessary condition to have a region of large momentum where a blue tilt dominates. But a blue tilt seems to be natural if we somehow change the sign of the effective scalaron mass i.e. if we convert, in the corresponding high energy region the scalaron in a normal form of matter contributing to the cosmological tilt with the opposite sign creating a blue tilted phase. There are some models, for instance, the curvaton axion model that achieve this goal. In any case the important message is that the primordial black hole solution to the dark matter problem requires already a form of matter changing at large energy scales the tilt from red to blue.

What we want to suggest, and we have worked our in detail in Ref.~\cite{GJ5} is that the blue tilted phase is the natural quantum source of the dark matter in the sense that i) contributes to the tilt as ordinary matter and ii) triggers the formation of primordial black holes that can appear to us as the relic of such pre-inflationary phase.

\subsection{Two phenomenological comments}
We conclude this section pointing out two potentially observable consequences of the quantum cosmological tilt. The first, is that the blue tilt for value of $k \eta > 1$ could yield the dark matter in the form of primordial black holes. This can be tested in the near future with upcoming experiments.

A feature that can be already tested with current observations is the behavior of the quantum cosmological tilt at scales ($\sim 10$ Mpc) of about $100$ times smaller than the current horizon. In this region, the small bump gives us a red tilt of order ($1-n_s \sim 0.1-0.3$) deviating significantly from scale invariance in the sense of generating less power on these scales. These scales can be tested observationally independently of the CMB.

The level of the power spectrum is usually represented by the combination of the root mean square of the fluctuations in a comoving top-hat filter of 8 Mpc and the current matter density ($\Omega_m$) as $S_8 = \sigma_8 (\Omega_m/0.3)^{0.5}$). The fluctuations can be measured by using the CMB determined value and extrapolating it to low redshift adopting the $\Lambda$CDM model. Alternatively, they can be measured directly from the low redshift universe by computing, for example, the correlation function of the weak lensing shear field. Extrapolating the $\Lambda$CDM model parameters, as measured by the Planck18 mission at the CMB, yields a value of $S_8 = 0.83 \pm 0.013$~\cite{Planck18} while measurements of the weak lensing correlation function by the KiDS team~\cite{KiDS} at redshifts of $z \sim 1$ yield to $S_8 = 0.759 \pm 0.022$. There is a discrepancy of about $3\sigma$ between the two values, with the present Universe being less clustered than the early one.

\begin{figure}
\centering
\includegraphics[width=.8\columnwidth]{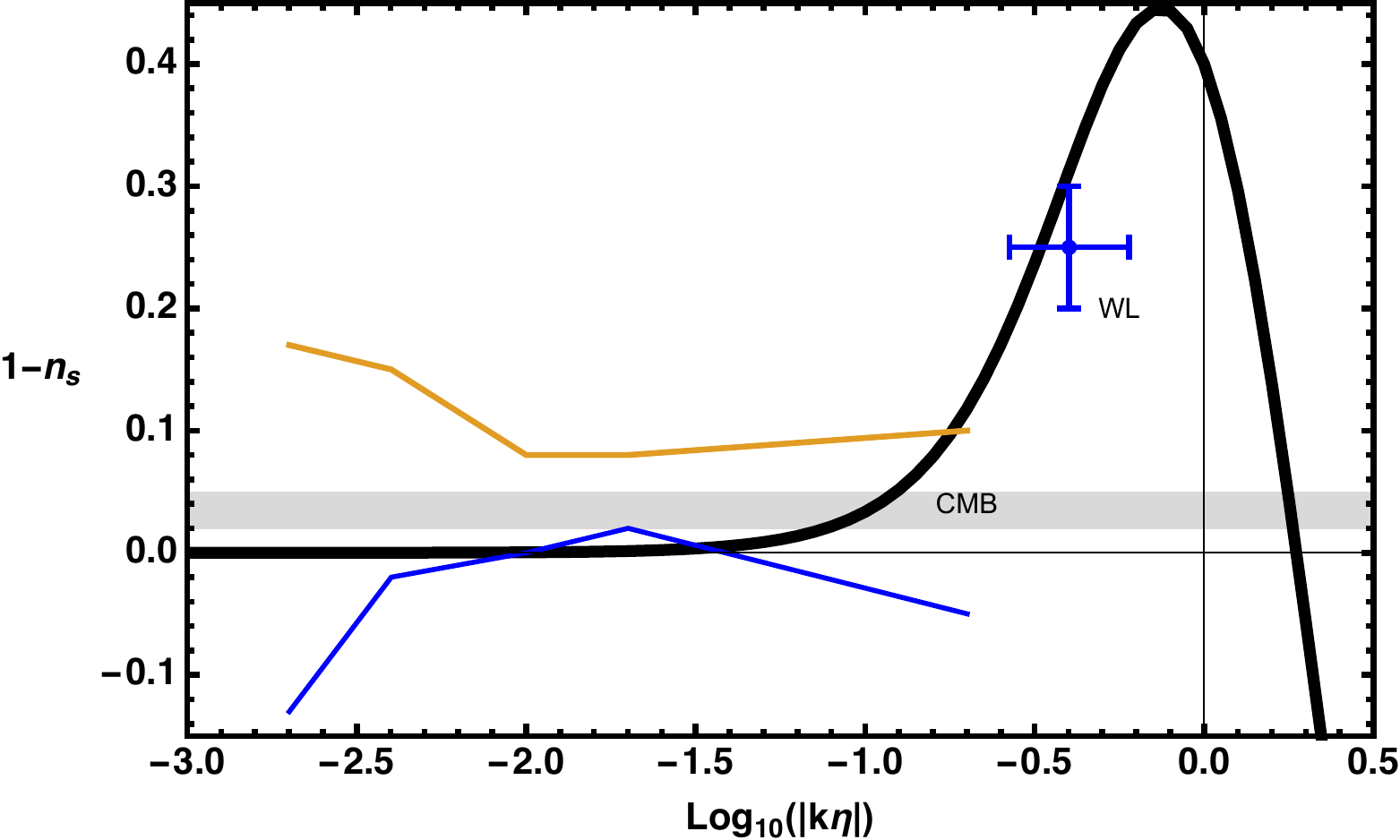} 
\caption{The black line shows the tilt in the power spectrum of primordial fluctuations as a function of scale computed from the quantum Fisher. Note that for scales a factor ten smaller than the current horizon, $1-n_s \sim 0.03$. Overplotted as orange and blue lines are the best constraints, at the 95\% level, on the reconstruction of $1-n_s$ from Planck data and large scale structure from Ref.~\cite{Ravenni} allowing the tilt to be a free function modeled by splines; this allows for the most freedom in the reconstruction. The blue dot and associated uncertainty bars shows the effective value of $1-n_s$ as derived from the local universe by the KidS-1000 weak lensing survey~\cite{KiDS}. This is a low redshift value and independent of the CMB.}
\label{fig:wl}
\end{figure}

The value for $1-n_s$ as inferred from the $S_8$ measurement of KiDS  is shown as the blue dot in Fig.~\ref{fig:wl} with corresponding uncertainties. This lack of power seems to fit well with the predictions of the quantum Fisher tilt. We also show the spline model independent reconstruction of $1-n_s$ by Ref.~\cite{Ravenni} (their Fig.~4), where we show the 95\% confidence contours. This is a reconstruction for larger scales and agrees well with our  the model prediction. In addition, the gray shaded area shows the value of $1-n_s$ assuming its value at all scales is that of the CMB.

%\appendix
\section*{Appendix: IR sensibility of the quantum cosmological tilt}

\begin{figure}
\centering
\includegraphics[width=.8\columnwidth]{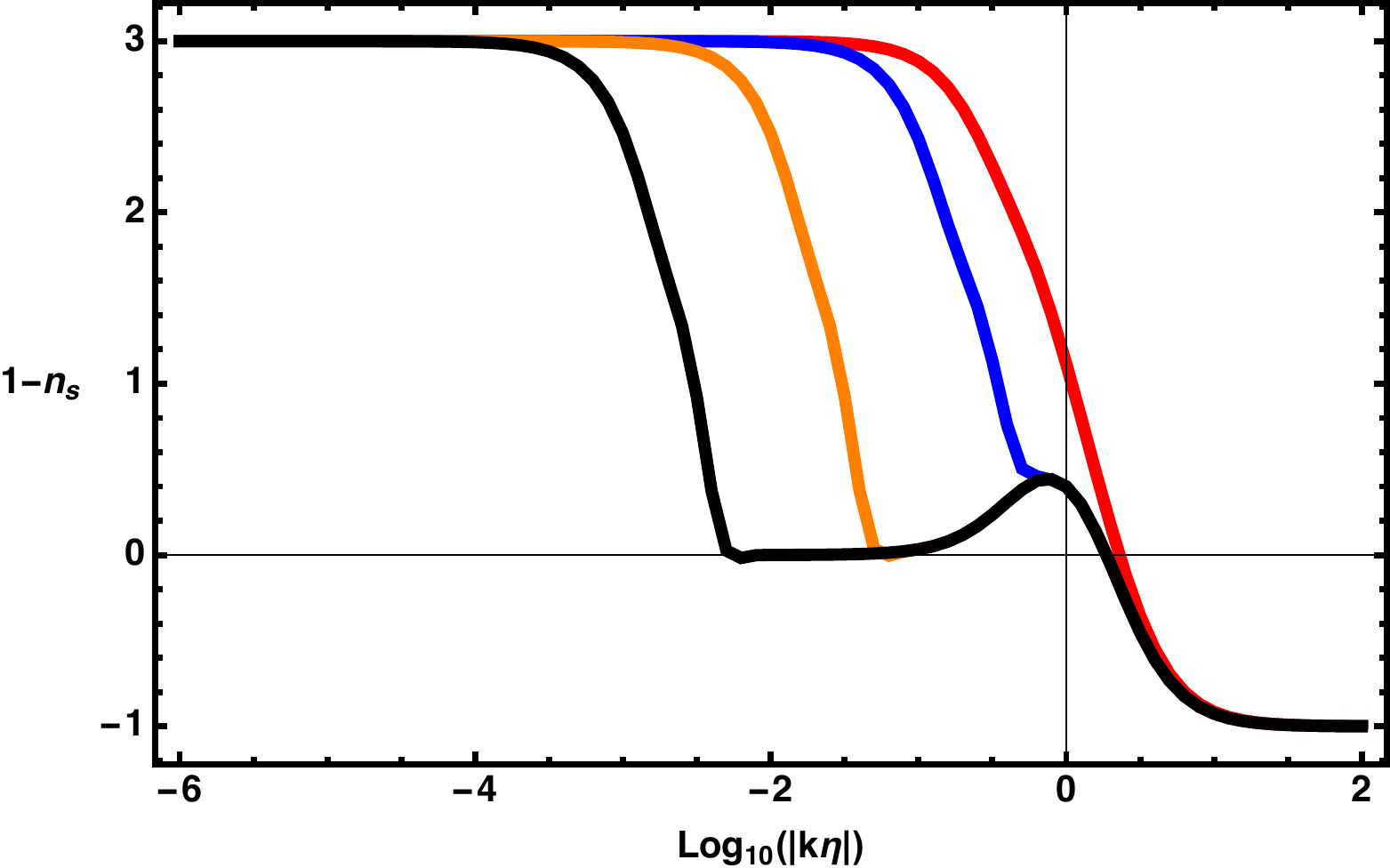} 
\caption{The effect of the number of pair production $n$ on the cosmological tilt $1-n_s$. The red, blue, orange and black curves are for $n=1, 10, 1000, 100,000$ respectively. The nearly scale invariant tilt is produced by the increase in $n$. See text for more details on how to set $n$ and end inflation.}
\label{fig:pairs}
\end{figure}

\begin{figure}
\centering
\includegraphics[width=.8\columnwidth]{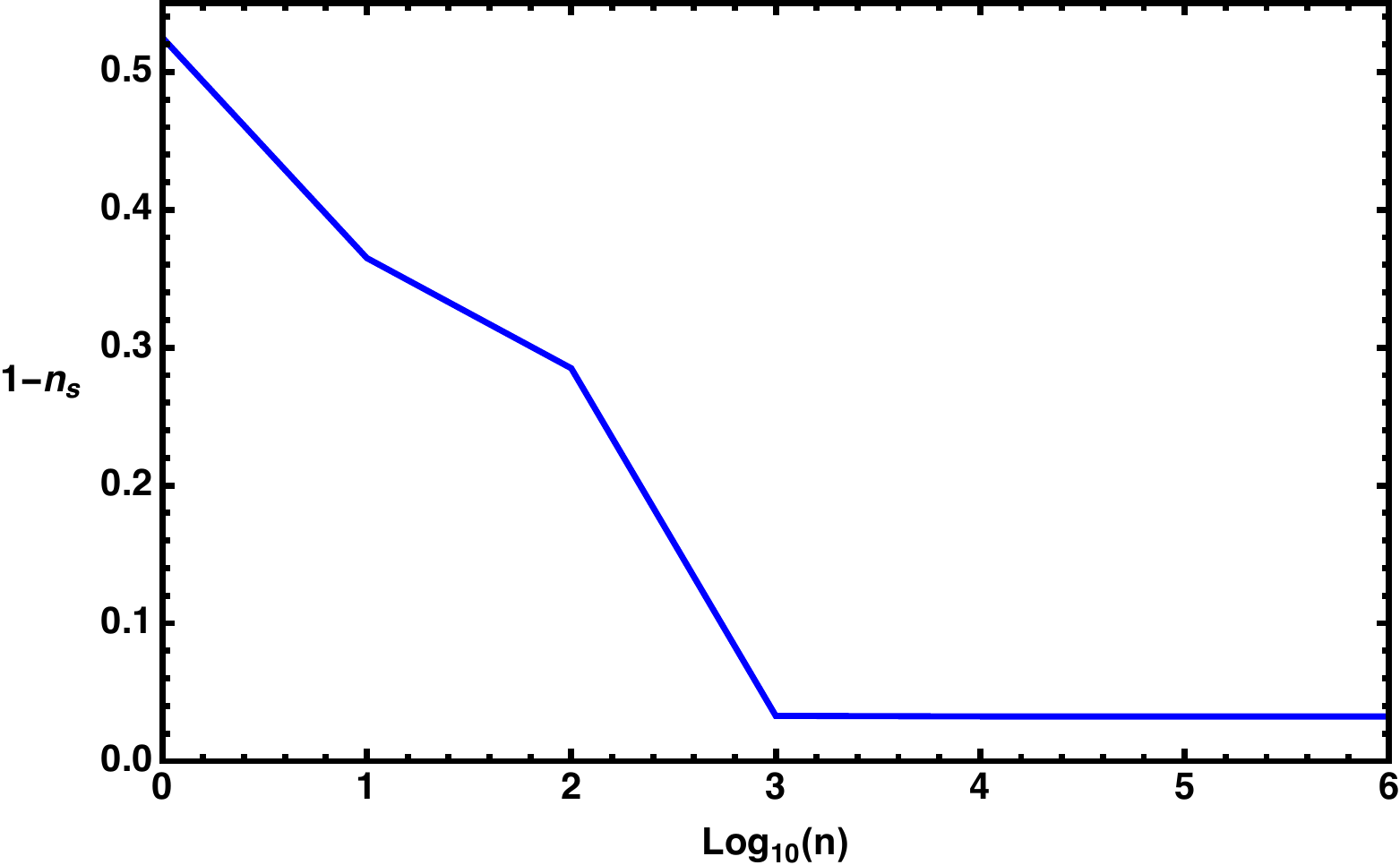} 
\caption{The value of the spectral index $1-n_s$ as a function of pair production $n$. For $n > 1000$ it converges to $1-n_s=0.0328$ and remains at that value.}
\label{fig:conver}
\end{figure}

The quantum Fisher information for the state $|\Lambda\rangle$ has been computed in the basis of  entangled pairs dictated by the de Sitter Bogolioubov transformation. Numerically, we have performed the sum over the number of pairs until a certain cutoff value $N$. In this section we discuss how the shape of the quantum cosmological tilt $\alpha_F$ depends on this cutoff. First of all, note that in the high energy regime of large values of $k\eta$, the contribution of states with large number of pairs is very suppressed, i.e. the quantum states in that region are coherent and not squeezed. Contrary, in the IR regime, with small $k\eta$ and large squeezing parameter, states with arbitrarily large value of soft entangled pairs contribute. In this sense, the UV regime of the quantum cosmological tilt should not be very dependent on the value of the cutoff $N$ while the IR regime should be very sensible to this cutoff. This phenomena is explicitly shown in Fig.~\ref{fig:pairs} where we observe two interconnected facts. On one side, we observe that the blue tilted UV phase is not changing when we increase the value of $N$. On the other side, we also observe that the almost scale invariant region becomes larger when we increase the value of $N$. Physically this effect of $N$ is teaching us two main lessons. First, that the almost scale invariant region we associate with the inflationary period appears as a consequence of the contribution of large numbers of entangled pairs. Until what value of the scale $k\eta$ extends this region is a consequence of the state normalization. The normalization factor depends on the squeezing parameter as $\frac{1}{cosh(r(k\eta))}$ with $r(k\eta) = \log(\sqrt{1+ \frac{1}{4(k\eta)^2}}+ \frac{1}{2k\eta})$. Thus in the deep IR region this normalization factor should depend on the number $N$ of pairs as $\frac{1}{\sqrt{N}}$. This sets the smaller value of $k\eta$ for the cutoff $N$ as
\begin{equation}\label{pairs}
    N\sim \frac{1}{(k\eta)^2}
    \end{equation}
In other words, for a given value $N$ the computed quantum cosmological tilt is reliable only until this IR scale. A priori we do not have any obvious physical criteria to set this cutoff on $N$. What we observe is that for a given large value of $N$ the extension in time of the almost scale invariant inflationary period can be estimated using (\ref{pairs}) to be
\begin{equation}
{\cal{N}} \sim \frac{1}{2}\ln N
\end{equation}
with ${\cal{N}}$ the number of e-foldings during the inflationary period. In other words, the end of inflation is set by the IR cutoff on $N$. In the numerical analysis we have performed we get some hints on how the system with an IR cutoff $N$ behaves when we formally push $(k\eta)$ to super horizon scales much smaller than (\ref{pairs}). In this regime, where we push $k\eta$ beyond the bound (\ref{pairs})  the system tendency, triggered by the normalization condition, is to leave inflation creating a large red tilt formally implying an epsilon larger than one. This is an appealing feature that is however out of the effective control of our computation so it should be taken with a grain of salt.

Finally, let us comment on the prediction of $(1-n_s)$ in CMB scales. This prediction was done solving the master equation and obviously depends on the form of $\alpha_F$ so we could wonder if this prediction is sensible to the IR cutoff $N$. In Fig.~\ref{fig:conver} we show that the value of $1-n_s$ converges for $n > 1000$ and remains at a value $1-n_s=0.0328$ for any value of $n > 1000$. In a certain sense the spectral index is a property of the effective {\it universality class} defined by taking the limit $N=\infty$ in the IR cutoff. In summary this value of the spectral index is very insensitive to the IR cutoff $N$.

So summarizing:
\begin{itemize}
\item The sum over large number of entangled pairs is crucial to have an almost scale invariant inflationary region.
    \item The UV blue tilted preinflationary region is a robust prediction that is not affected by the IR cutoff on $N$.
    \item The predicted value of the spectral index using the master equation is equally robust and not affected by the IR cutoff.
    \item The IR cutoff $N$ sets formally the duration of the inflationary period.
\end{itemize}
As a very speculative final comment we could try to use some entropy argument to set a bound on the value of $N$. If we interpret $\ln N$ as an entropy measure we could bound $N$ by requiring  $\ln N$ to be smaller than $N_{GH}$ the Gibbons Hawking de Sitter entropy. We will investigate quantitatively further this superficial comment in the future.

\acknowledgments
 The work of CG was supported by grants SEV-2016-0597, FPA2015-65480-P and PGC2018-095976-B-C21. The work of RJ is supported by MINECO grant PGC2018-098866-B-I00 FEDER, UE. RJ acknowledges ``Center of Excellence Maria de Maeztu 2020-2023" award to the ICCUB (CEX2019- 000918-M).

\end{document}